\documentclass[prb,twocolumn,preprintnumbers,amsmath,amssymb]{revtex4}

\usepackage{graphicx}
\usepackage{dcolumn}
\usepackage{bm}

\begin{document}

\title{Evidence for a large magnetic heat current in insulating cuprates}

\author{M. Hofmann$^1$, T. Lorenz$^1$, K. Berggold$^1$, M. Gr\"{u}ninger$^1$,
        A. Freimuth$^1$, G.S. Uhrig$^2$, and E. Br\"{u}ck$^3$}

\address{$^1$II. Physikalisches Institut, Universit\"{a}t zu K\"{o}ln,
         50937 K\"{o}ln, Germany\\
         $^2$Institut f\"{u}r Theoretische Physik, Universit\"{a}t zu K\"{o}ln,
         50937 K\"{o}ln, Germany\\
         $^3$Van der Waals-Zeemann Laboratorium, University of Amsterdam,
         1018 XE Amsterdam, The Netherlands}
\date{\today}

\begin{abstract}
The in-plane thermal conductivity $k$ of the two-dimensional
antiferromagnetic monolayer cuprate Sr$_2$CuO$_2$Cl$_2$ is studied.
Analysis of the unusual temperature dependence of $k$ reveals that at low
temperatures the heat is carried by phonons, whereas at high temperatures
magnetic excitations contribute significantly. Comparison with other
insulating layered cuprates suggests that a large magnetic contribution
to the thermal conductivity is an intrinsic property of these materials.
\end{abstract}

\pacs{PACS numbers: 66.70.+f; 74.72.-h; 75.10.jm}

\maketitle

There is growing experimental evidence that spin excitations may
contribute significantly to the heat current in low-dimensional spin
systems. This seems to be well established for one-dimensional (1D)
systems \cite{ando98,sologubenko00,sologubenko01,hess01,lorenz02}. For
example, in the insulating spin-ladder material
Sr$_{14-x}$Ca$_x$Cu$_{24}$O$_{41}$ a large magnetic contribution $k_m$ to
the thermal conductivity $k$ can be derived from a pronounced double-peak
structure of $k$ along the ladder direction \cite{sologubenko00,hess01}.
The situation is less clear in two-dimensional (2D) spin systems. These
are, however, of particular importance due to their relevance for
high-temperature superconductivity \cite{johnston97,kastner98}. A
double-peak structure comparable to that in 1D systems is found in the
in-plane thermal conductivity of insulating 2D cuprates such as
La$_2$CuO$_4$ (LCO) and YBa$_2$Cu$_3$O$_6$ (YBCO) (Ref.\
\onlinecite{nakam91,cohn95,hessthesis}). This may indicate a sizable
magnetic contribution to the heat current at high temperatures
\cite{nakam91}. However, the phononic thermal conductivity $k_{ph}$ may
show a double-peak structure also, as a result of pronounced (resonant)
scattering in a narrow temperature range. Such scattering may arise from
the presence of local magnetic excitations, as was recently shown for the
2D spin-dimer system SrCu$_2$(BO$_3$)$_2$ (Ref.\ \onlinecite{hofmann01}),
or it may arise from the presence of soft phonon modes \cite{cohn95}. The
latter was suggested for LCO and YBCO, in which soft modes e.g.
associated with tilt distortions of the CuO polyhedra are known to be
present \cite{cohn95,latticeinstability}. An additional complication
arises from a strong sensitivity of the double-peak structure to light
oxygen doping \cite{cohn95}.

A material of particular interest in this context is Sr$_2$CuO$_2$Cl$_2$
(SCOC). It is structurally very similar to LCO: It contains $\rm
CuO_2$-layers as in LCO, but the out-of-plane oxygen ions at the apices
of the CuO$_6$-octahedra are replaced by Cl and La by Sr.  The material
has several advantages compared to LCO and YBCO (see e.g. Ref.\
\onlinecite{johnston97}): (1) SCOC does not exhibit any distortion from
tetragonal symmetry down to at least 10K so that there is no structural
instability associated with soft tilting modes. (2) Because of the
absence of tilt distortions the magnetic properties are simpler than
those of LCO. For example, there is no Dzyaloshinski-Moriya exchange
interaction. Thus, SCOC is believed to represent the best realization of
a two-dimensional square-lattice $S=1/2$ Heisenberg antiferromagnet. (3)
In contrast to LCO and YBCO, SCOC cannot be doped easily with charge
carriers.

In this paper we present measurements of the in-plane thermal
conductivity $k$ of SCOC. We identify a double-peak structure from a
pronounced high-temperature shoulder around 230 K. Analysis of these data
and a comparison to LCO and YBCO shows that it is very unlikely that the
double peak structure arises from anomalous phonon damping due to
scattering on soft lattice modes or magnetic excitations. The data
indicate instead a large magnetic thermal conductivity at high
temperatures as an intrinsic feature of the insulating 2D cuprates.

We studied a single crystal of Sr$_2$CuO$_2$Cl$_2$ of rectangular form
($1 \times 3 \times 4$ mm$^3$) with the short direction along the
crystallographic $c$ axis. It was grown by the traveling-solvent floating
zone method. The thermal conductivity was measured with the heat current
within the CuO$_2$~planes by a conventional steady-state method using a
differential Chromel-Au+0.07\%Fe-thermocouple. Typical temperature
gradients were of the order of 0.2~K. The absolute accuracy of our data
is restricted by uncertainties in the sample geometry whereas the
relative accuracy is of the order of a few~$\%$.\cite{surfacelayer}.

\begin{figure}
\begin{center}
     \includegraphics[width=0.9\columnwidth,clip]{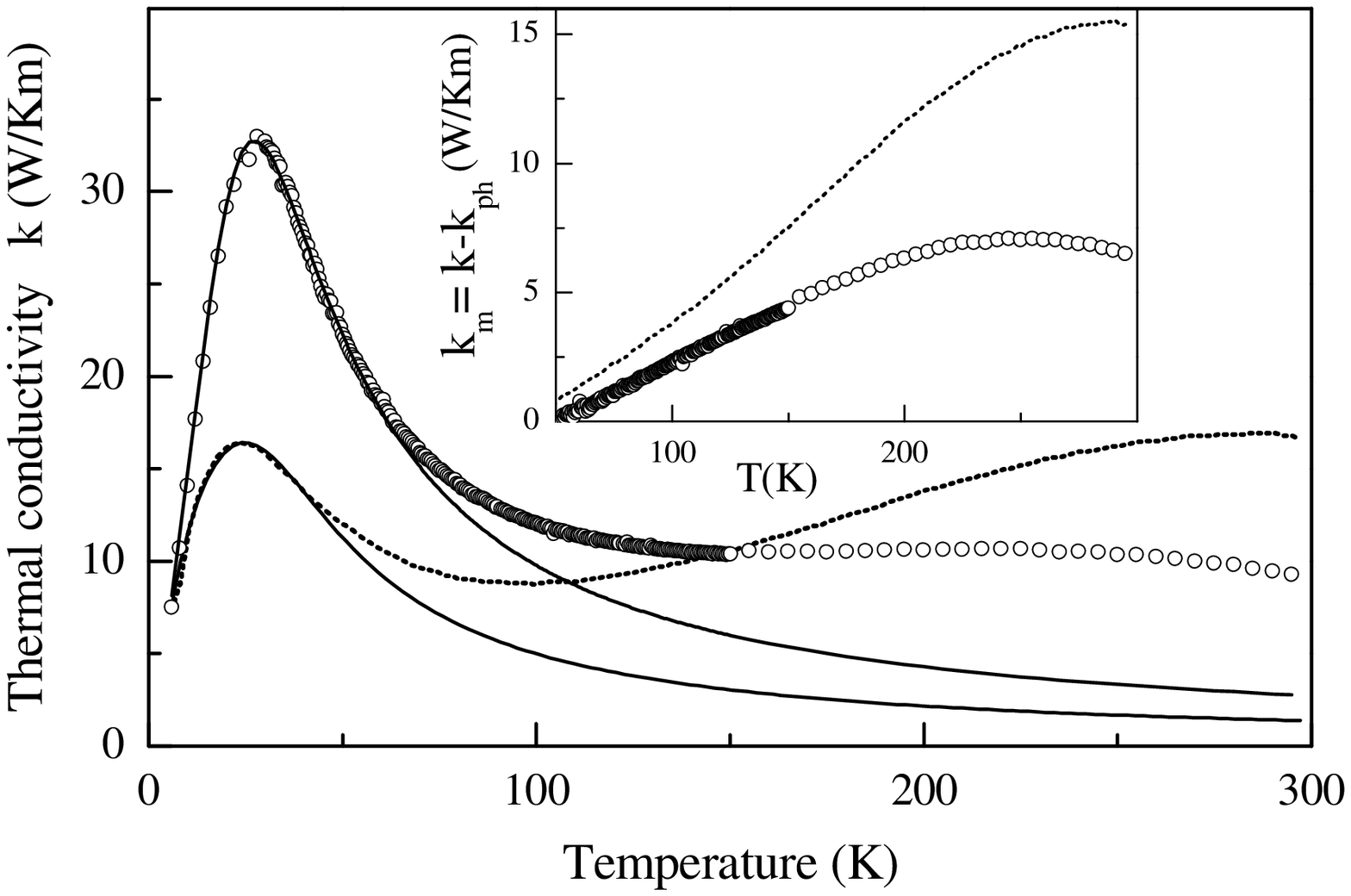}
\end{center}
\caption{In-plane thermal conductivity $k(T)$ of Sr$_2$CuO$_2$Cl$_2$
(circles) and La$_2$CuO$_4$ (dotted line; data from Nakamura et al.
\cite{nakam91}). Solid lines: fits to $k_{ph}$ using the Debye model
\cite{hofmann01,berma76}. The fitting parameters for Sr$_2$CuO$_2$Cl$_2$
(La$_2$CuO$_4$) are: $D/10^{-17}$s = 3.4 (2.6); $P/10^{-43}$s$^3$ = 0.15
(21); $U/10^{-30}$s$^2$/K = 1.0 (2.2); $u$ = 4.9 (4.4). The point defect
scattering ($P$) is smaller in SCOC. Inset: $k_m = k-k_{ph}$ for
Sr$_2$CuO$_2$Cl$_2$ (circles) and La$_2$CuO$_4$ (dotted line) (see
text).} \label{fig1}
\end{figure}

We show in Fig.~\ref{fig1} the in-plane thermal conductivity of SCOC as a
function of temperature. We identify a maximum at $\approx$~30~K and a
shoulder at high temperatures around 230K. The pronounced low-temperature
maximum of $k$ indicates a high crystal quality. We note that $k$ is
independent of a magnetic field ($\le 8$ Tesla) applied within the
CuO$_2$~planes perpendicular to the heat current. For comparison we show
in Fig.~\ref{fig1} the in-plane thermal conductivity of a single crystal
of LCO measured by Nakamura et al.~\cite{nakam91}. These data also reveal
a double-peak structure. The absolute value of $k$ at the low-temperature
maximum is smaller than in SCOC. One reason may be that LCO is more
sensitive to defects, resulting e.g.~from excess oxygen, which introduces
lattice defects and hole doping and thus reduces the mean free path of
the heat-carrying excitations.

Both compounds, SCOC and LCO, are antiferromagnetic insulators. In an
insulator the heat is usually carried by phonons. The typical behavior of
$k_{ph}$ of a crystalline insulator is shown by the solid lines in
Fig.~\ref{fig1}. These curves represent fits to the low-temperature
maximum of $k$ (fitted below about 50 K) of SCOC and LCO using the
standard Debye model for the thermal conductivity of acoustic phonons
\cite{hofmann01,berma76}:
\begin{equation}
k_{ph} = \frac{k_{B}^4 T^3}{2 {\pi}^2 \hbar^3
  v_{ph}}
\int\limits_{0}^{\Theta_{D}/T}
 \tau\left(x,T\right) \frac{x^4 e^x}{\left( e^x - 1 \right)^2}dx \ .
\label{kphonon}
\end{equation}
Here $\Theta_{D}$ is the Debye temperature and $v_{ph}$ the sound
velocity. Due to the lack of experimental data for SCOC we use for both
compounds the values reported for LCO ($\Theta_D \approx 385$~K
\cite{junodginsberg}; $v_{ph}$ $\approx 5.2\cdot 10^3$~m/s
\cite{suzuki00}). $\omega$ is the phonon frequency, $x =
\hbar\omega/k_{B}T$, and $\tau\left(x,T\right)$ is the phonon relaxation
time given by
\begin{eqnarray}
\nonumber
{\tau}^{-1} =  \frac{v_{ph}}{L} + D \omega^2 + P \omega^4 + U T \omega^3 \exp
\left(\frac{\Theta_D}{uT} \right) \, .
\label{rates}
\end{eqnarray}
The four terms refer to the scattering rates for boundary scattering,
scattering on planar defects, on point defects, and phonon-phonon Umklapp
scattering, respectively. $L \approx 1 $mm is the sample length, and $D$,
$P$, $U$, and $u$ are fitting constants. The low-temperature data are
described very well by these fits. The fit parameters are given in the
caption of Fig.~\ref{fig1}. The decrease of $k_{ph}$ at high temperatures
is due to phonon-phonon Umklapp scattering.

For several reasons it is very unlikely that the high-temperature
increase of $k$ is due to conventional heat transport by phonons: (1) The
contribution to $k$ from acoustic phonons, as described above, decreases
at high temperatures. (2) The contribution of optical phonons to the heat
current is usually much smaller than that of acoustic phonons, even in
compounds with a very large number of atoms in the unit cell
\cite{lorenz02,slack79}, so that heat transport by optical phonons is
very unlikely to cause the high-temperature maximum. (3) The out-of-plane
thermal conductivity $k_c$ of LCO behaves as the in-plane thermal
conductivity $k$ at low temperatures, but $k_c$ shows no indication of a
high-temperature maximum \cite{nakam91}. Such strongly temperature
dependent anisotropy is not expected for purely phononic heat conduction.

Additional phonon scattering, active in a narrow temperature range close
to the minimum of $k$, may in principle cause a double-peak structure.
However, resonant scattering on local magnetic excitations as in
SrCu$_2$(BO$_3$)$_2$ (Ref.\ \onlinecite{hofmann01}) cannot be the correct
explanation in the present case: In the 2D square-lattice cuprates the
dispersion of magnetic excitations ranges from $\approx 0$ to $2 J/k_B
\gtrsim 2000$ K ($J$ is the in-plane exchange constant) so that there is
no reason that scattering on magnetic excitations should be most
pronounced in a narrow temperature interval around 100
K.\cite{hofmann01,sales02} Note, in particular, that in the 2D cuprates
scattering on magnetic excitations should not disappear above the
N\'{e}el-temperature $T_N$, because the relevant energy scale is set by $J\gg
T_N$ (see below). Additional phonon damping from scattering on soft
lattice modes as suggested in Ref.\ \onlinecite{cohn95} is also unlikely
as a cause of the double peak: (1) There are no lattice instabilities in
SCOC, rendering this mechanism unimportant for this material. (2) The
double-peak structure is also present in the tetragonal low-temperature
phase of Eu-doped LCO, in which no soft tilting modes should be present
either \cite{hessthesis}. (3) The absence of a double-peak structure of
$k_c$ in LCO (Ref.\ \onlinecite{nakam91}) implies that anomalous phonon
scattering would have to be active only for $k$. Such strong anisotropy
of the phonon-phonon scattering is not expected. Finally, note that the
finding $k_c < k$ in LCO (Ref.\ \onlinecite{nakam91}) provides evidence
against any scattering scenario as a cause of the double peak structure:
for such scattering, if active only for $k$, but absent for $k_c$, implies
$k<k_c$, in contradiction to the experimental results.

The data of Fig.~\ref{fig1} (in particular $k > k_c$) are most naturally
explained, if an additional channel of heat transport for the in-plane
thermal conductivity is present. In an undoped insulating 2D Heisenberg
antiferromagnet with an electronic gap $\gtrsim 1.5$ eV, the only
candidate for heat transport next to phonons are magnetic excitations.
Their thermal conductivity $k_m$ adds to that of the phonons, i.e.\ $k =
k_{ph} + k_m$. In order to extract $k_m$ from the data we subtract
$k_{ph}$ as obtained from the fit of the low-temperature maximum. Note
that $k_m$ cannot be obtained at $T \lesssim 100$K in this way, because
Eq.~\ref{kphonon} was fitted to the {\it total} $k$ below 50~K.
Remarkably, $k_m$ is of comparable magnitude (roughly of the order 10
W/Km) in both compounds (see inset Fig.~\ref{fig1}). The maximum of $k_m$
is at $\approx $ 245 K in SCOC and at $\approx $ 285 K in LCO.

\begin{figure}
\begin{center}
\includegraphics[width=0.9\columnwidth,angle=0,clip]{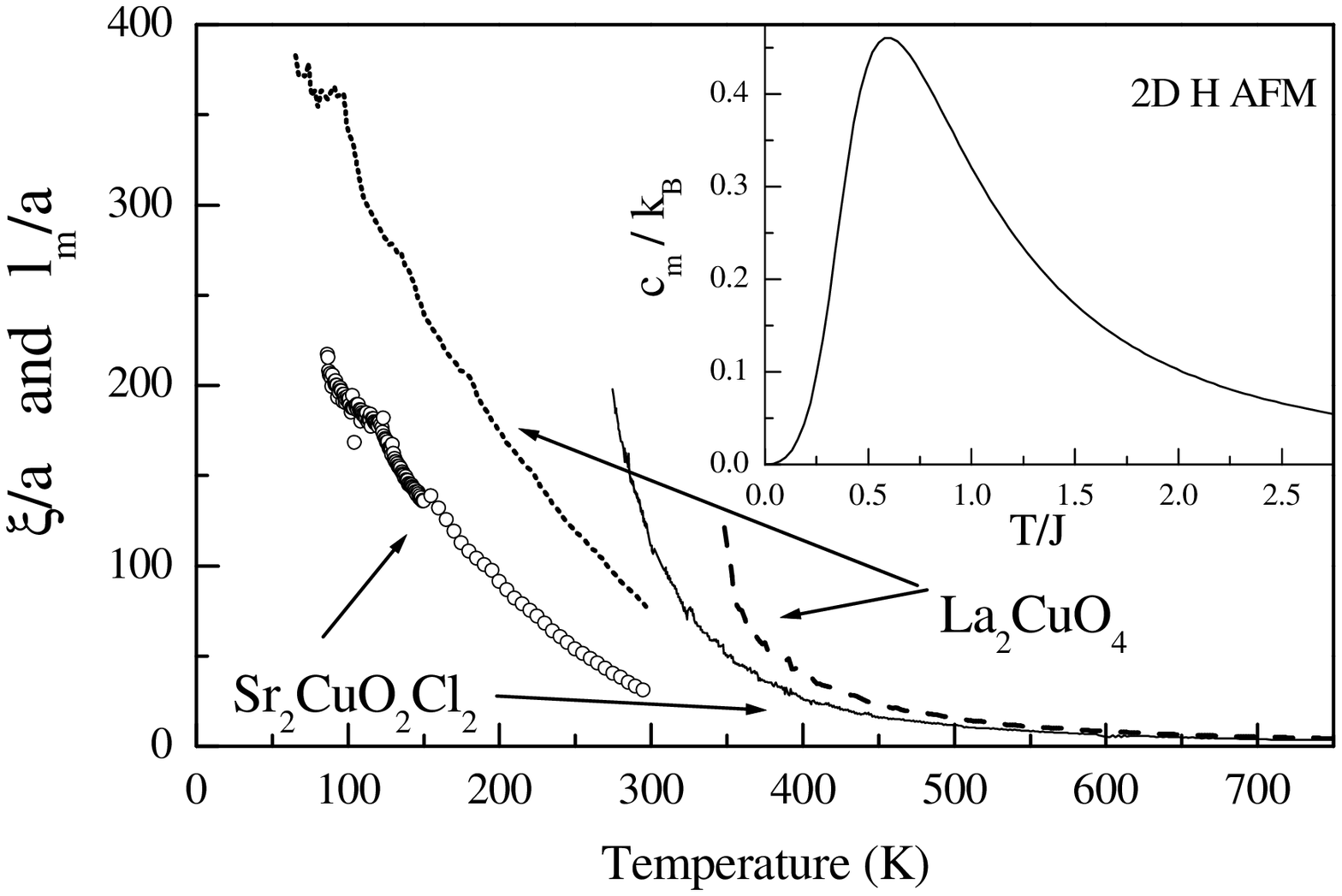}
\end{center}
\caption{Magnetic mean free path $\ell_m(T)/a$ (circles:
Sr$_2$CuO$_2$Cl$_2$; dotted line: La$_2$CuO$_4$) as extracted from the
data shown in Fig.~\ref{fig1} (see text), and in-plane magnetic
correlation length $\xi_m(T)/a$ (solid line: Sr$_2$CuO$_2$Cl$_2$; dashed
line: La$_2$CuO$_4$) as obtained from neutron scattering
\cite{kastner98}. Here, $a\simeq 3.9$ \AA is the lattice constant. Inset:
Specific heat $c_m/k_B$ of a $S=1/2$ square-lattice Heisenberg
antiferromagnet (see appendix).} \label{fig2}
\end{figure}

Is a magnetic contribution of this size reasonable? We estimate $k_m$
using the kinetic equation in 2D \cite{berma76}:
\begin{equation}
k_m = \frac{1}{2} c_m v_m \ell_m \, .
\label{kinetic}
\end{equation}
Here $c_m$ is the magnetic specific heat and $\ell_m$ the mean free path
of the magnetic excitations. The velocity $v_m$ of long-wavelength spin
waves is $v_m^{\rm SCOC} \approx 1.06 \cdot10^5$ m/s and $v_m^{\rm LCO}
\approx 1.16 \cdot10^5$ m/s, which is obtained from $v_m = \sqrt{8}S Z_c
J a/\hbar$. Here $Z_c$ is the Oguchi correction.\cite{oguchi} Values of
$J$ for various 2D cuprates are given in Table I. Note that $v_m$ is much
larger than $v_{ph}$, as a result of $J \gg k_B \Theta_D$. For the
specific heat $c_m$ we use the theoretical result shown in the inset of
Fig.~\ref{fig2}, which comes from the extrapolation of the
high-temperature series for the partition sum (see appendix). The maximum
is $c_{\rm max}=0.4612(5) Nk_B$ at $k_BT_{\rm max}=0.5956(1)J$. Given
$v_m$, $c_m$, and using $k_m$ as shown in Fig.~\ref{fig1} we obtain an
estimate of $\ell_m(T)$ using Eq.\ (\ref{kinetic}). $\ell_m$ decreases
strongly with increasing temperature (Fig.~\ref{fig2}). At room
temperature, $\ell_m/a \approx 30$ for SCOC and $\approx 75$ for LCO.
These values are not excessively large -- much larger values of $\ell_m$
have been found in one-dimensional spin systems
\cite{sologubenko00,sologubenko01,hess01} -- rendering a magnetic
contribution to the heat current in SCOC and LCO very plausible.

For a better understanding of $k_m$ it is instructive to discuss the
magnetic correlations in the quasi-2D cuprates, in particular, the
in-plane magnetic correlation length $\xi_m(T)$. In a 2D antiferromagnet
long-range order with $\xi_m = \infty $ is restricted to $T$=0. With
increasing temperature spin-flips (or magnons) are excited, which reduce
$\xi_m$ by breaking the long-range correlation. In the quasi-2D materials
considered here, the finite magnetic ordering temperature $T_N$ is
determined by the {\em inter-}plane interaction
\cite{johnston97,kastner98}, which is much weaker than the in-plane
exchange interaction $J$, so that $k_B T_N \ll J$ (see Tab. I). In the
ordered state at $T<T_N$, $\xi_m = \infty$. For $T>T_N$, $\xi_m $ is
still large because of the large $J$. We show $\xi_m(T)$ of SCOC and LCO
as inferred from magnetic neutron scattering \cite{kastner98} in
Fig.~\ref{fig2}. Above $T_N$, $\xi_m$ is indeed much larger than the
lattice constant. However, $\xi_m$ decreases strongly with increasing
temperature, approximately according to \cite{kastner98} $\xi_m(T) \simeq
\exp (2 \pi J/k_B T)$.

From these considerations we may draw several conclusions for the
magnetic heat current: (1) We expect that $k_m$ is determined by the
large in-plane exchange interaction $J$ and not by the inter-plane
interactions. Therefore, no significant anomaly of $k_m$ is expected at
$T_N$. (2) At least above $T_N$ the heat-carrying magnetic excitations
are not the familiar collective excitations of a magnetically ordered
state (i.e.\ conventional magnons), but rather magnetic excitations (of
triplet character) in a spin-liquid state. Note, however, that also at
$T<T_N$ the nature of the magnetic excitations of the 2D cuprates is
under intensive debate
\cite{grueninger00,ho01,aeppli99,sandvik01,laughlin97}. (3) The strong,
exponential decrease of $\xi_m$ above $T_N$ suggests a similar decrease
of $\ell_m(T)$. This would explain, why the maximum of $k_m$ occurs at a
temperature much lower than that of $c_m$: the increase of
$c_m(T<T_{max})$ is overcompensated by the strong decrease of
$\ell_m(T)$. Note, however, that the results on
Sr$_{14-x}$Ca$_x$Cu$_{24}$O$_{41}$ show that in a spin-liquid $\ell_m$
may be significantly larger than $\xi_m$.\cite{sologubenko00,hess01}

\begin{table}
\begin{tabular}{lcccc}
   & $T_H$ (K)  & $T_N $ (K) & $J/k_B$ (K) & $J/k_B T_H$\\ \hline
YBa$_2$Cu$_3$O$_6$    & 200        & $>400$ & 1125    &  5.6   \\
Sr$_2$CuO$_2$Cl$_2$   & 245        & 260   & 1220    &  5.0  \\
La$_2$CuO$_4$         & 285        & 320   & 1390    &  4.9
\end{tabular}
\caption{Position $T_H$ of the high-temperature maximum of $k_m$,
Neel-temperature $T_N$, in-plane magnetic exchange coupling constant $J$,
and the ratio $J/k_B T_H$ for three insulating 2D cuprates. Note that
$J/k_B T_H$ is very similar in all three compounds. $T_N$ is from Refs.
\protect\onlinecite{johnston97,kastner98,rossat}. The values of $J$ are
derived from two-magnon Raman scattering and infrared
bimagnon-plus-phonon absorption data \cite{lorenzana,grueninger00}, where
the Oguchi correction has been taken into account.\cite{oguchi} The data
for $k$ in YBCO and LCO are from Refs. \protect\onlinecite{cohn95} and
\protect\onlinecite{nakam91}, respectively. } \label{tab1}
\end{table}

A double-peak structure of $k$ has also been reported for the insulating
bilayer compound YBa$_2$Cu$_3$O$_6$.\cite{cohn95} As in LCO one observes
pronounced anisotropy, i.e.\ $k_c$ does not show a high-temperature
maximum. Given the existence of a large $k_m$ in the monolayer cuprates
LCO and SCOC, the high-temperature maximum of YBCO is also likely to be
of magnetic origin. This is consistent with the systematic variation of
the temperature $T_H$ of the high-temperature maximum of $k$ with $J$ for
the three different insulating cuprates as shown in Tab.\ \ref{tab1}.

The data in Fig.~\ref{fig1} show that the high-temperature maximum is more
pronounced in LCO than in SCOC. A related observation is the weak
high-temperature anomaly of $k$ in Pr$_2$CuO$_4$.\cite{Sologubenko99b} In
view of their rather similar magnetic properties one would expect a
similar behavior of $k_m$ in these compounds.\cite{DMinteraction}
However, doping with mobile charge carriers \cite{cohn95,hessthesis} or
static impurities \cite{taldenkov96,hessthesis} influences the
high-temperature maximum strongly. In particular, the high-temperature
maximum in YBCO depends in a non-monotonic way on the oxygen
concentration \cite{cohn95}. Thus the different behavior of $k_m$ might
result from the fact that LCO and YBCO can easily be doped with charge
carriers via variation of the oxygen content, whereas this is not
possible in SCOC. Further experiments with a detailed control of the
charge-carrier concentration could clarify this issue.

In summary, the in-plane thermal conductivity of Sr$_2$CuO$_2$Cl$_2$ shows
an unusual temperature dependence with a pronounced shoulder at high
temperatures, similar to the behavior found for La$_2$CuO$_4$
\cite{nakam91} and YBa$_2$Cu$_3$O$_6$.\cite{cohn95} There is no
structural instability in Sr$_2$CuO$_2$Cl$_2$. Moreover, scattering on
magnetic excitations is not restricted to a narrow temperature interval
around the minimum of $k$ in these compounds. It is therefore unlikely
that the double peak structure arises from strong damping of the phononic
heat current by resonant scattering on soft lattice modes or magnetic
excitations. The data rather indicate a large magnetic contribution to
the heat current as an intrinsic property of the 2D cuprates.

We acknowledge useful discussions with M. Braden, W. Hardy, A.P. Kampf,
V. Kataev, D.I. Khomskii, and G.A. Sawatzky. This work was supported by
the Deutsche Forschungsgemeinschaft in SFB 608 and in SP 1073.
A.F.~acknowledges support by the VolkswagenStiftung.

\begin{center} {\bf APPENDIX} \end{center}


To deduce $c(T)$ from the high-temperature series \cite{U2} this series is
converted in a series for the entropy $s$ in the energy per site $e$
(Ref.\ \onlinecite{U3}). The extrapolations are stabilized by information
on the ground state energy $e_0=-0.669437(5)$ (Ref.\ \onlinecite{U4}),
the maximum entropy $s=\ln 2$, and the  expected low-energy power law
$s(e) \propto (e-e_0)^{2/3}$. For the latter, Pad\'e approximants are
applied to $s'(e)/s(e)-2/(3(e-e_0))$ (Dlog Pad\'e approximation). Very
good results are obtained (reliable error estimate $10^{-2}$ from
comparing diagonal to non-diagonal Pad\'e approximants). The result shown
in the inset of Fig.~\ref{fig2} (error $10^{-3}$) is obtained by
approximating
\begin{equation}
\left[ \frac{(e-e_0)s'(e)}{s(e)} - \frac{2}{3} \right]
\ln\left(\frac{e-e_0}{1-e_0}\right) \, ,
\end{equation}
which allows for multiplicative logarithmic corrections, yielding $c(T)
\propto T^2(A+\ln^{-\gamma}(1/T))$ with $A=0$ and $\gamma=1.05(5)$. We do
not, however, exclude a small finite value of $A$ as found in spin-wave
theory.


\begin{references}
\bibitem{surfacelayer}
To obtain reproducible data the samples should not be exposed to air,
because SCOC is strongly hygroscopic.
\bibitem{DMinteraction}
We do not expect a pronounced influence of the Dzyaloshinski-Moriya
interaction present in LCO on $k_m$.
\end{references}
\end{document}